\documentclass[a4paper,fleqn]{cas-dc}

\usepackage[numbers]{natbib}

\def\tsc#1{\csdef{#1}{\textsc{\lowercase{#1}}\xspace}}
\tsc{WGM}
\tsc{QE}
\tsc{EP}
\tsc{PMS}
\tsc{BEC}
\tsc{DE}

\begin{document}
\let\WriteBookmarks\relax
\def\floatpagepagefraction{1}
\def\textpagefraction{.001}
\shorttitle{}
\shortauthors{}

\title [mode = title]{Attention-Wrapped Hierarchical BLSTMs for DDI Extraction}   

\tnotetext[1]{This document is the results of the research
   project funded by the National Science Foundation.}

\author[1]{Vahab Mostafapour}[type=author,
                        orcid=https:0000-0003-4414-5706
                        ]
\ead{vahabmspour@gmail.com}

\author[1]{Oguz Dikenelli}[type=author,
                        orcid=https:0000-0001-7948-7453]
\ead{oguz.dikenelli@ege.edu.tr}

\address[1]{Department of Computer Engineering, Ege University, Izmir, Turkey}

\begin{abstract}
Drug-Drug Interactions (DDIs) Extraction refers to the efforts to generate hand-made or automatic tools to extract embedded information from text and literature in the biomedical domain. 

Because of restrictions in hand-made efforts and their lower speed, Machine-Learning, or Deep-Learning approaches have become more popular for extracting DDIs. In this study, we propose a novel and generic Deep-Learning model which wraps Hierarchical Bidirectional LSTMs with two Attention Mechanisms that outperforms state-of-the-art models for DDIs Extraction, based on the DDIExtraction-2013 corpora. This model has obtained the macro F1-score of 0.785, and the precision of 0.80.
\end{abstract}




\maketitle
\section{Introduction}

Drug-Drug Interactions (DDIs) usually refers to changes in the effects of a drug which occur as results of the presence of another drug. Although these changes are some-times useful, the result may be harmful when the interaction increases the toxicity of a drug \cite{baxter2010stockley}.
Hazardous effects, along with high costs for patients and insurance companies, have turned the discovery of drug interactions into an important, and widespread scientific research area. The results of these researches are generally reported in medical journals, and literature  \cite{aronson2007communicating}.

Drug-Drug Interaction (DDI) Extraction problem is described as the application of Information Extraction (IE) techniques to biomedical literature  in order to discover embedded interactions between drugs. 

There are some published databases and resources such as Drugbank \cite{knox2010drugbank}, and Stockley's Drug Interactions \cite{baxter2010stockley}, have gathered and categorized a number of Drug-Drug Interactions, and their types. However, there are currently two major problems with these resources:  they are mainly made by health care professionals, and are usually updated only once every two years. Therefore, the development of an automatic extraction tool would accelerate and simplify the browsing of medical publications and literature to extract certain detected interactions. 

SemEval-2013-task-9 challenge \cite{segura2013semeval}, and its corpus \cite{herrero2013ddi}, boosted the application of Machine Learning (ML) techniques on DDI extraction problem. 

Former generations of ML techniques were mainly based on statistical methods such as Support Vector Machine (SVM) \cite{kim2015extracting,  segura2011using, bell2011statistical, bui2014novel}, while recent approaches have moved toward  the use of Deep Neural Networks (known as Deep-Learning) architectures, which have been shown to perform better than statistical models \cite{quan2016multichannel, Liu2016, zhao2016drug, zhang2017drug, lamurias2019bo}.
These methods have transferred some effective features such as word features, POS tags, Dependency Graphs, and Parse Trees from statistical methods, and use them as the input for networks. Additionally, the inputs or features in deep-learning methods are generally mapped into a low dimensional vector space, such as a predefined embedding representation vectors using Mikolov Word2Vec \cite{mikolov2013distributed}, GloVe \cite{pennington2014glove}, or BioBERT \cite{lee2019biobert}, or into a randomly generated representation.

In this study, we propose the \textbf{Attention-Wrapped Hierarchical BLSTMs (AW-BLSTMs)} model for DDIs Extraction. 
The main objective of AW-BLSTMs is to use the lowest possible level of external information and without big cleaning and pruning operations on the input data. This means that we aimed to provide a model that is robust for real-life scenarios even with uncleaned data and be easily applicable to different languages even those have not strong syntactic passers.

The novelty of AW-BLSTMs model is that it utilizes the two levels of attention mechanisms that wrap two layers of BLSTMs to tackle DDI extraction problem, and could outperform state-of-the-art DDIs extraction models.

The first level attention mechanism, called \textit{entity-level}, is applied to assign a weight to each word based on its contextual relatedness to the target entities that comes from their embedding space. From this mechanism, the model can highlight the words that are more related to the target entities. The top level attention mechanism is applied on the top of BLSTM layers, before the softmax classifier. This mechanism, which feeds from the output of BLSTM, gives the model the opportunity to learn the weight for the sentences-wide features that are 
related to the annotated relationship.

This model outperforms  the classification metrics of CNN and RNN based state-of-the-art DDIs extraction models (in our knowledge) in both the  F1-score (0.785) and the   precision score (0.80) in the macro-average approach.

\section{Related Works}
Deep Neural Networks (DNNs) such as Recurrent Neural Network (RNN) and Convolutional Neural Network(CNN)  models have shown a good performance in the DDI Extraction task.

One of the successful CNN based models to extract DDIs is presented in \cite{Liu2016}, which leverages word and position embedding. They obtained a 0.698 for the F1-score on the DDI-Extraction-2013 corpora. A multi-channel CNNs model is presented by \cite{quan2016multichannel}, which used five different types of word embeddings to get the best embedding performance possible. Another remarkable CNN model, used to detect DDIs from biomedical literature, is the Syntax Convolutional Neural Network (SCNN) which could reach a 0.686 in the F1-score based on the DDIExtraction-2013 corpora \cite{zhao2016drug}. (Asada et al.) \cite{asada2017extracting} have presented a CNNs based model for DDI extraction task that is intensified with an Attention mechanism which could help the model to obtain a 0.6912 F1-score.

Although the CNNs have pointed out a good performance in DDIs Extraction, RNNs like Long Short-Term Memory (LSTM) networks have managed to outperform them in this task  in recent years \cite{zhang2017drug, zheng2017attention, sahu2018drug}. Besides the LSTM networks, applying Attention Mechanisms to the DNNs is another successful approach to increase the performance of DDIs or general domain semantic relationship Extractors. This mechanisms give to their models the opportunity to highlight the effect of the features with high relatedness to the potential relationships in the input sentence.  

(Sahu et al.) \cite{sahu2018drug} has presented three LSTM based models for the DDI extraction task that are called B-LSTM, AB-LSTM, and Joint AB-LSTM. These models have employed LSTMs, and attention mechanisms and achieved a 0.6939 F1-score as their highest value. They are designed to categorize the potential relationship into five classification types: \textit{"Advice"}, \textit{"Effect"}, \textit{"Mechanism"}, \textit{"Int"}, \textit{"Negative"} (instead of Non-type relationship). \cite{zhang2017drug} has introduced a hierarchical RNNs (combination of LSTM and BLSTM) model to extract DDIs from entity annotated medical text. In this model, an attention mechanism is applied to the input layer to take advantage of the closeness of each word to the target entities. It also shows that the usage SDP can significantly improve the F1-score that has obtained 0.729. Because the usage of SDP increases the dependence of the model on language, we do not use it in our model.
\cite{zheng2017attention} has published a RNNs based model for  DDIs extraction with focus on data cleaning operations. The architecture of their model is similar to \cite{zhang2017drug} model, but a big effort on input data pruning, cleaning and removing redundant sentences has helped them outperform the other models with F1-score 0.773, which is the highest value in state-of-are models.

BO-LSTM model \cite{lamurias2019bo}, has leveraged a domain ontology knowledge to intensify the model that sound so beneficial and achieved 0.75 percent F1-score that is higher than most of previous works. 

Our model takes advantage of BLSTM networks as its intermediate layers in a hierarchical form. The first layer is used for capturing the information of each separated input, and the second layer captures information of whole inputs together. Some previous works such as \cite{zhang2017drug} and \cite{zheng2017attention} have used an entity attention mechanism alongside LSTM layers. Some models, such as \cite{sahu2018drug}, have leveraged an attention layer on the top of their LSTM layer. However, this is the first time that  a model that is presented for DDIs extraction,  combines different attention mechanisms (input-attention and output-attention) with BLSTM layers, so that the attention layers wrap the hierarchical BLSTM layers and outperformed the state-of-the-art models. This combination helps the model to highlight the more effective words of the given sentence and selected features on the extraction performance together. 

\section{Model Architecture and Materials}

In this section, we describe the architecture of the  \textbf{AW-BLSTMs} model. The main contribution of this model is to employ two types of attention layers to surround a hierarchical BLSTMs to tackle the DDI extraction problem as described above. These attention layers are used to modify the input and output of BLSTMs layers by different attention mechanisms then outperform the standard BLSTMs. In the following sections, a detailed description is given.

\subsection{ Input Features and Embeddings}

\textbf{\hspace{0.3 cm}Input Features:}  Various methods are used to exploit more features and information from the biomedical text in order to outperform DDIs extractors, or any biomedical relationship extractor in general. These works vary from rule-based methods to deep-learning methods \cite{hao2005discovering, segura2011linguistic, segura2011using, zhang2017drug}.

\begin{figure}
\includegraphics[width=1.1\linewidth]{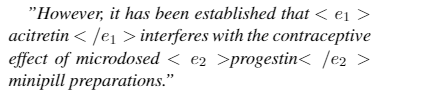}
\caption{\textbf{Sample Sentence      }}
\label{Sample-Sentence}
\end{figure}

\begin{figure}
\includegraphics[width=1.1\linewidth]{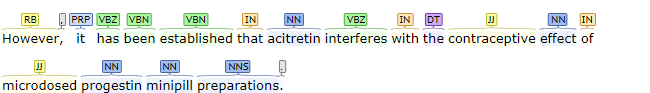}
\caption{\textbf{Part-of-Speech tag for the sample sentence}}
\label{fig-POS}
\end{figure}

The most recent deep-learning studies on DDIs extraction have boosted their model by some strong features of the input sentences such as Words, POS tags, and Positions (related distances) \cite{nguyen2018convolutional, zhang2017drug, zeng2014relation}. The words of the sentences carry essential information about relationships between target entities, but t the effects words vary due to their distance to each target entity. \cite{zeng2014relation} has addressed the influence of the embedded information in the positions of the target entities within the sentence. That is, a network can obtain more information about target drugs from the words that are closer to them than those that are farther away which can cause noise in the model. To access this information, the position feature of words is often used.  In addition, it is argued that the Part Of Speech (POS) tags information fortifies the input layer and relation extraction model to improve its performance. Moreover, the influence of these features are also compared in some other DDIs extraction models such as \cite{zhang2017drug}. The SDP feature is escaped to keep model language independent and generic enough to be applicable to the other domains and languages. So in this work Words, POS tags and Position (relative distances) features are employed as input features.
Figure~\ref{Sample-Sentence} presents an example sentence with two annotated target entities.  Figure~\ref{fig-POS} shows the Part-of-Speech tags of the sample sentence~\ref{Sample-Sentence} that are taken from Stanford-Core-NLP tool \cite{manning-EtAl:2014:P14-5}. In this example, the token \textit{"interferes"} is tagged by \textit{VBZ} as its POS tag which indicates that it is a form of a verb. 

\textbf{Embeddings:} Words embedding refers to the vector representation of a vocabulary. It maps each word of the vocabulary to a vector in a real space. In fact, it is a transformation of vocabulary to a high-dimensional one-hot vector space (with the dimension of vocabulary size), or to a low dimension space which can be obtained by randomly initialized vectors or training a neural network such as Mikolove Word2Vec \cite{mikolov2013distributed}, GloVe \cite{pennington2014glove}, BioBERT \cite{lee2019biobert}, etc. 

Embedding representation has played a successful role in neural networks based information extraction models in the NPL domain \cite{palangi2016deep, zeng2014relation,zhang2017drug}. It enables models to grabs context and semantic information of words of a document. In this work, an embedding representation for word and POS tags is trained by Word2Vec \cite{mikolov2013distributed} on the PubMed abstracts who includes `drug` trigger word. Also, the standard normal distribution is employed to initialize an embedding space for position feature \cite{wang2016relation}, \cite{zhang2017drug}.

\begin{figure}
\centering
\includegraphics[width=1.1\linewidth]{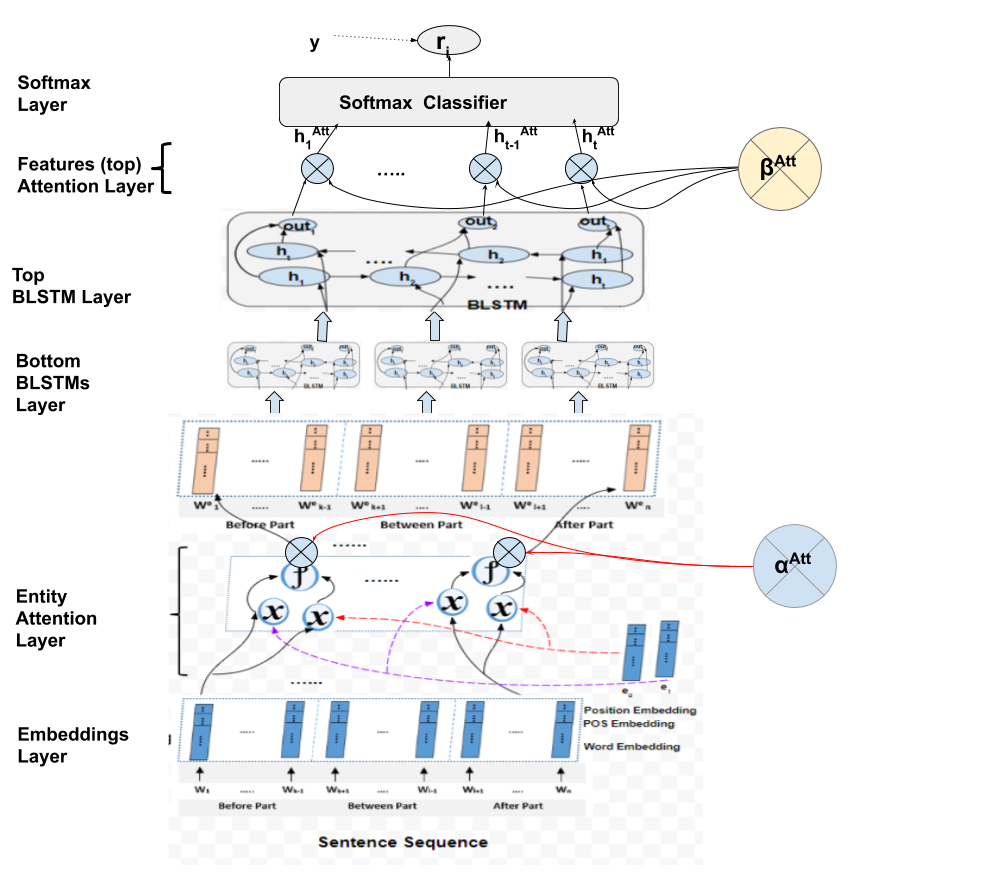}
\caption{Attention-Wrapped Hierarchical BLSTMs (AW-BLSTMs) model for DDIs Extraction}
\label{fig:AW-BLSTMs}
\end{figure}

\subsection{Model Layers}

Figure~\ref{fig:AW-BLSTMs} illustrates the general architecture of our model including its attentions and BLSTMs layers.

\subsubsection{Input Layer}
\begin{figure}
\centering
\includegraphics[width=1.1\linewidth]{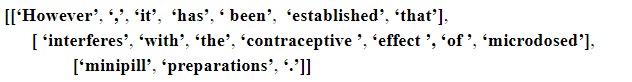}
\caption{Partitions of a Sentence}
\label{fig:sentence-parts}
\end{figure}

\textbf{\hspace{0.3 cm} Words-Sequence:} We partition each sentence into three parts; (i) Before part (tokens come before the first mentioned entity), (ii) Between part (tokens between two mentioned entities) and (iii) After part (tokens come after second mentioned entity). Figure~\ref{fig:sentence-parts} illustrates this partitioning strategy for words of the example sentence in Figure~\ref{Sample-Sentence}. This strategy helps the network to learn their effect separately in its lower layers. It also solves the problem of long sentences that are removed by \cite{zheng2017attention} in their data cleaning steps. 
Words of each sentence are indicated by an index to utilize the embedding look-up table in the embedding layer.

\textbf{POS-Sequence:} Part-of-Speech tags that gathered from Stanford-Core-NLP tool \cite{manning-EtAl:2014:P14-5}. Parallel to the Words-Sequence, POS tags Sequence is partitioned into three parts as well. 

\textbf{Position-Sequence:} Position sequences or distance sequences are the relative distance of each word related to position of both target entities that are presented by $dist_{e_1}$, $dist_{e_2}$ respectively. For example,  in the given example sentences Figure~\ref{Sample-Sentence}, $dist_{acitretin}(\text{"effect"} ) = 5$, $dist_{progestin}(\text{"effect"} ) = -2$. So $dist(\text{"effect"} ) = [5, -2]$.  Similar to POS tags sequence and word sequence, positions sequence is partitioned into three parts too.

\subsubsection{Embedding Layer}
Embedding layer encodes the input sequences $S$ of each word $W_i$ to a low-dimensional vector space. Each word $W_i$ includes three features Word, POS tag and position (which is a pairs of related distances to both target entities). Suppose that $WV_i$ is the embedding representation of word $W_i$  with its features. Then each $WV_i$ is a concatenation of word-embedding vector $wv_i$ (embedding representation of word $W_i$ and trained on PubMed abstracts), POS-embedding vector $pos_i$ (also trained on PubMed abstract), and position-embedding vector $[dist_{e_1}, dist_{e_2}]$ (mapped each distance value to a 10-dimensional vector initialized by normal random distribution). Then the input sentence (or a sub-sentence; one part) S can be displayed as a list of embedding-vectors as follows: 
\begin{equation}\label{input-sen}
S  = [WV_0,WV_1,..,WV_{maxLen}]
\end{equation}

That is, for each word $W_i$ includes its three features, we have the following concatenated embedding vector:
\begin{equation}
WV_i = [wv_i, pos_i, [dist_{e_1}, dist_{e_2}]] 
\end{equation}
Such that, 

$wv_i \in R^{WE_{dim}}$ ; $WE_{dim}=$ dimension of word-embedding.

$pos_i \in R^{pos_{dim}}$ ; $pos_{dim}=$ dimension of POS-tag-embedding.

$dist_{e_j} \in R^{dist_{dim}}$,  j = 1, 2; $dist_{e_j}$= distance of word $wv_i$ related to $e_j$th target entity.

$dist_{dim}=$ dimension of position-embedding or distances embedding; 

\label{sect-3.2.2}

\subsubsection{Entity Attention Layer}

In long and complicated sentences, and in those having multiple clauses, there will likely be some words and clauses that could misguide the model to a different relation label than the desired one. Assigning weight to the words, is a successfully solution that has been applied to overcome such problems \cite{wang2016relation}, \cite{zhang2017drug}. This method is called the \textit{Attention Mechanism}, which allows the models to learn the weight of each word's contribution on the target interaction. The essence of this mechanism is a vector ($\alpha$ vector) with the size of the sentence length. It is considered to keep a weight for each word. In our model, this mechanism is applied in two different levels: (i) in the entity level (Input level attention) and (ii) the top level (feature level). The first is used to highlight more effective words of the sentence in terms of two target units. The latter one is applied over the top of the bidirectional RNNs and gives the capability to the model to precisely learn the contribution rate of the features. A detailed explanation comes as follows:

\textbf{Entity Attention:} In this study, entity level attention takes advantage of our embedding over PubMed training corpus since the silent semantic relationships between words of the corpora are reflecting in the unsupervised trained vectors by  Mikolove Word2vec \cite{mikolov2013distributed}. To turn this idea into account, the weight  vector $\alpha$  is computed as follows:

\begin{equation}
    \alpha_j^i = \frac{exp(f(e_i, wv_j))}{\sum_{k}{exp(f(e_i, wv_k))}}
\end{equation}

$f(e_i,wv_j)$ is the function that computes semantic relatedness of each word $wv_j$ with target entity $e_i$ which in this work it is computed as the inner product of embedding vectors of $e_i$ and $wv_j$. 

In the next step to benefit from the joint impact of $\alpha{}_j^1$ and $\alpha{}_j^2$ (attention vectors related to target entity $e_1$ and $e_2$ respectively) together, a simple average of them is considered as follows.

\begin{equation}
\alpha{}_j^{Att}=\frac{\alpha{}_j^1 + \alpha{}_j^2}{2}
\end{equation}
Finally by applying attention vector on the embedding vector of each word in sentence sequence, embedding vector of each word updates due to the following  function:
\begin{equation}
  WV_j^{Att} = dot(\alpha{}_j^{Att},WV_j) 
\end{equation}
After applying thıs attention mechanism layer on the sentences, they feed the bottom bidirectional LSTM cells. 

\subsubsection{BLSTM Layers}

Recurrent Neural Networks (RNNs) have played an important role in Information Extraction and NLP tasks in recent years. The characteristic that makes RRNs more appropriate to apply on text mining and NLP problems is their sequential structure. This structure provides the capability to learn the dependencies between the inputs by connecting previous information (state) to the present. The most commonly used, and successful version of RNNs are Long Short-Term Memories (LSTMs) that are a modified RNNs which introduced by \cite{hochreiter1997long} to solve the long term dependencies and Gradient Vanishing problem. LSTMs units benefits from structures called cell state and gates. The gates provide the capability of removing and adding information to the cell state,  which then serves as an information belt over time-steps. Each unit updates in  time-step t due to following operations:
\begin{equation}
f_t=\sigma(W_f .[h_{t-1},x_t]+b_f)
\end{equation}
\begin{equation}
i_t=\sigma(W_i .[h_{t-1},x_t]+b_i)
\end{equation}
\begin{equation}
g_t=tanh(W_g .[h_{t-1},x_t]+b_g)
\end{equation}
\begin{equation}
C_t=(f_t \otimes C_{t-1}) \oplus (i_t \otimes g_t)
\end{equation}

Such that $\sigma$ is the sigmoid activation function and the operations $\otimes$,  $\oplus$ indicate the point-wise tensors multiplication and summation respectively.  

\textbf{BLSTMs:}  The information flows of the input sentence are from left to right in an LSTM network. So, the network is only able to capture the one-side dependencies and relationships between tokens. To allow the network to keep the effect of backward relationships between tokens (from the end of the input sequence toward its begin) another LSTM layer with the reverse direction (backward-side) is usually applied which is called Bidirectional LSTM or BLSTM in brief. The output of BLSTMs is the concatenation of these two layers.
Formally, suppose an input sequence $S$ like equation~\ref{input-sen}: 
the forward LSTM takes an element of S in each step from $WV_0$ to $WV_{maxLen}$ respectively, and generates its outputs
$H^\rightarrow=[h^\rightarrow_1, h^\rightarrow_2, ... , h^\rightarrow_n]$. In contrast, backward LSTM takes the input $S$  reversely from $WV_{maxLen}$ to $WV_0$ respectively to calculate its outputs $H^\leftarrow=[h^\leftarrow_1, h^\leftarrow_2, ... , h^\leftarrow_n]$ and $n$ is the number of LSTM units and $T$ is the number of time-steps or sequences length. Thus, the BLSTM turns out the features vector $H = [ H^\rightarrow, H^\leftarrow]$.

In our work, as displayed in Figure ~\ref{fig:AW-BLSTMs}, BLSTM layer consists of two layers. The first BLSTMs layer is applied on the entity attention layer. In this layer, three BLSTMs(with same configurations) are placed on top of each part (Before, Between, and After parts) of the input sequence to capture information embedded in each part independently. This lower BLSTMs turn out $H_{1_{Befor}}$, $H_{1_{Middle}}$.,and $H_{1_{After}}$.The outputs of this layer, $H_1$s, are fed into the higher BLSTM layer. The second BLSTM layer is applied to the top of the lower  BLSTMs. So, the higher BLSTM layer lies on the whole input sentence and connects separated parts together. that is, it takes outputs of feature vectors of all three BLSTMs as its inputs then generates its output vector $H_2$, 
\begin{equation}\label{H2}
H_2=[h_2^1, h_2^2, . . . , h_2^l]
\end{equation}

This layer allows our model to learn the features that rely on all parts of input together.

\subsubsection{Top Attention Layer}

The second attention mechanism is applied to our model to give it the opportunity to concentrate on the sentence-wide features which are more effective in the relationship classification phase. This mechanism allows the model to learn weights for the sentence level syntactic and semantic features that are reflected in the output of the upper BLSTM layer. That is, these features convey the effects of all operations from the lowest level operation that is the embeddings attention, to the highest one, the second bidirectional LSTMs. The model highlights the vector representation of these features by assigning weight them. The model learns these weights during its training and keeps them in a vector called top attention vector or $ \beta$. Following equations show the Mathematical computation steps of this attention vector: 
\begin{equation}
\beta_j=dot(W_{top}, h_2^j)+ b_{top}
\end{equation}

s.t. $h_2^j$ is $j^{th}$ element of $H_2$ in the equation ~\ref{H2}

\begin{equation}
\beta_j=tanh(\beta_j)
\end{equation}
\begin{equation}
    \beta_j^{Att} = \frac{exp(dot(U_{top},\beta_j))}{\sum_{k}{exp(dot(U_{top},\beta_k))}}
\end{equation}

We have employed the simple $dot$ operation to apply the effect of attention vector $ \beta$ on the $H_2$ output vector and generate attention affected output $H_2^{Att}$. This vector is ready to feed the softmax classifier layer.

\begin{equation}
H_2^{Att}=dot(\beta^{Att}, H_2)
\end{equation}

\subsubsection{Classification Layer (Softmax layer)}
Attention-based weighted feature vectors, $H_2^{Att}$, are final outputs of the inner layers of our model. These vectors are fed into the \textit{softmax} classifier to predict the appropriate relation type. The softmax classifier is set in a dense or fully connected layer with the number of nodes equal by the number of class types (5, is this case) as its activation function. This function computes the conditional probability value for each input sentence using the $H_2^{Att}$ weighted vectors and predicts the highest value as the class number of the relation type \cite{zhou2016attention, wang2016relation}. That is, for DDI type y:
\begin{equation}\label{softmax}
p(y \mid x)= softmax(W_{dense}.H_2^{Att} + b_{dense})
\end{equation}

The softmax layer returns the class label y as its probability value as the input sentence x is the highest value.  We represent this label $r$ that is computed as follows:
\begin{equation}
r = argmax_y  p(y \mid x)
\end{equation}

For the training loss, the negative "log-likelihood" of the correct labels is leveraged as:
\begin{equation}
L = \sum_{x}{log(p_x)}; \hspace{0.4 cm} \text{x is the given sentence}
\end{equation}

\section{Result and Discussions}

\subsection{Dataset Description}

Following previous works and, in order to have a fair comparison, we have evaluated \textbf{AW-BLSTMs} model using the well-known DDIExtraction-2013 corpus \cite{segura2013semeval}. This corpus is a human annotated corpus, published in XML data format. Its Data sources are from Drugbank \cite{knox2010drugbank} text (792 texts), and Medline abstracts (233 abstracts). Among these files, there are 624 train files, and 191 test files, related to drug-drug interactions. This corpus consists of 5,021 sentence-level drug-drug interaction instances which are known as positive samples. It includes more than 28k negative samples that are tagged as false or not any desired relationship type, as well. To obtain train and test data, it is randomly split into about \%77  and  \%23 respectively. 

the objective of DDI Extraction problem due to the mentioned corpus is to classify the extracted relation between two annotated drugs into one of the \textit{"Advice"}, \textit{"Effect"}, \textit{"Mechanism"}, and \textit{"Int"} interaction types. We have adapted this problem to the multi-classification problem. To this end, negative samples (a pair of drug instances in the sentence which does not express any type of relationships) are labeled as \textit{"other"} type interaction. By this strategy, both detecting and classification sub-tasks of DDI Extraction task is satisfied.

\subsection{Evaluation Metrics and Performance comparison}
\textbf{Evaluation Metrics:} We evaluate the model by the most commonly used Precision, Recall, and  F1-scores measures. The F1-score is traditionally computed by
\begin{equation}
    F1-score =2. \frac{Precision.Recall}{Precision + Recall}
\end{equation}
which is harmonic mean of precision and recall.

\begin{table*}
\centering
\small
\begin{tabular}{cc}
\begin{tabular}{l|llll|lll}
\hline
&& \multicolumn{2}{c}{\textbf{F1-Score of DDI-Types}} & & & \multicolumn{2}{c}{\textbf{Overall}}\\\cline{2-5} \cline{6-8}
\textbf{Model} & \textbf{Advice } & \textbf{Effect} & \textbf{ Mechanism} &\textbf{ Int} &\textbf{Precision} &\textbf{ Recall} & \textbf{ F1-Score}\\\hline

B-LSTM \hspace{0.2 cm }\cite{sahu2018drug} & 0.7592 & 0.6515 & 0.7266 & 0.4740 & 0.6907 & 0.6435 & 0.6663 \\
BR-LSTM\hspace{0.2 cm }\cite{xu2018leveraging} & 0.7518 & 0.6817 & 0.7911 & 0.4336 & 0.7152 & 0.7079 & 0.7115 \\
BO-LSTM\hspace{0.2 cm }\cite{lamurias2019bo} &-&-&-&-&-&-& 0.751 \\
Joint AB-LSTM\hspace{0.1cm }\cite{sahu2018drug} & 0.8026 & 0.6546 &  0.7226 & 0.4411 &  0.7447 & 0.6496 & 0.6939   \\
HRNN\hspace{0.2 cm }\cite{zhang2017drug} & 0.803 & 0.718 & 0.740 & 0.543 & 0.741 & 0.718 & 0.729 \\
Att-BLSTM\hspace{0.2 cm }\cite{zheng2017attention}  &  \textbf{0.851} & 0.766 & 0.775 & 0.577 & 0.759 & 0.687 & 0.773 \\
RHCNN \hspace{0.2 cm }\cite{sun2019drug} & 0.8054 & 0.7349 & \textbf{0.7825} & \textbf{0.589} & 0.773 & 0.7375 & 0.7548 \\
AW-BLSTM & \textbf{0.819} & \textbf{0.774} & \textbf{0.78} & \textbf{0.584} & \textbf{0.80} & \textbf{0.77} & \textbf{0.785}   \\
\hspace{0.3 cm }(our model) \\\hline
\end{tabular} & 
\end{tabular}
\caption{Performance Comparison of AW-BLSTMs and other state-of-are deep
learning based ddi models }\label{tab:f1-scor-state-of-art}
\end{table*}

\textbf{Performance comparison:} Comparison of our model with other state-of-the-art Neural networks-based methods on DDIExtraction-2013 corpus is presented in this section. These results are reported in macro-averaged evaluation metrics for DDIs classification. The B-LSTM model presented by \cite{sahu2018drug} has been selected as the baseline model as contains only simple BLSTM layer the can show the effect of attention mechanisms. Table ~\ref{tab:f1-scor-state-of-art} presents the comparison of our model with some other RNN and CNN based state-of-art DDIs extraction models.

In the first part, one can see F1-score of models related to the four drug-drug interaction types separately.
Our model outperforms all in "Effect" type and shows comparable results in the rest of the types, so that in each type passes most of the other models except one (variously). 

In the second part,right side of the table, it shows the overall (in macro-averaged) results which our model outperforms them in all Precision, Recall and F1-score by 0.80, 0.77, 0.785 respectively. The best F1-score of the state-of-art model is presented in \cite{zheng2017attention} with F1-score 0.773, which is focused on data cleaning and removing redundant samples from gold standard. Although, the real world data is notoriously noisy and very often requires data prepossessing, as the they mentioned the steps are not applicable to all types data (the don't apply it on Madeline data) and also is not our purpose because of losing generality of the model to tackle real-world scenarios. However, the elimination of redundancy sentences is expected to improve the proposed model. We notice that the model presented in \cite{sahu2018drug} uses attention mechanism on the top of its BLSTMs layer but it could obtain at most 0.69 for F10 score and the model presented in \cite{zhang2017drug} has used entity level attention mechanism which could reach 0.717 for F1-score without SDP and 0.729 even with using SDP. Indeed using combined attention mechanisms that could capture information of relatedness words of with target entities together with information of the features lied on the sentences can intensify the model to the learn weights more appropriately.

\begin{table}
\centering
\small
\begin{tabular}{cc}
\begin{tabular}{llll}
\hline
\textbf{DDI-type} & \textbf{Precision} & \textbf{Recall} & \textbf{F1-score}\\\hline
Effect & 0.734  & 0.819  & 0.774 \\
Mechanism & 0.818  & 0.745  & 0.78 \\
Advice & 0.772  & 0.873  & 0.819 \\
Int & 0.776  & 0.469  & 0.584 \\
Negative & 0.968  & 0.967  & 0.968 \\\hline
\end{tabular} & 
\end{tabular}
\caption{Evaluation results of each DDI-types using of AW-BLSTMs model}\label{tab:all-rell-types}
\end{table}

In Table ~\ref{tab:all-rell-types}, we show the evaluation results of four drug-drug interaction types existing in the corpus called \textit{"Advice"}, \textit{"Effect"}, \textit{"Mechanism"}, and \textit{"Int"}. The maximum Precision is 0.818 \textit{"Mechanism"} and  maximum F1-score value is for \textit{"Advice"} with 0.819. Also, we can see that the values of "Negative" is higher than 0.96.

\subsection{Implementation and Experimental Setup}
Implementation operations took place on the keras library with Tensorflow Backend. 
To train the model, code is executed in Jupyter Notebook on Anaconda environment. Code runs on a PC with core i5 CUP and 12GB RAM.
The best result of F1-score is reached on epoch 101 with batch-size=64 and validation-split=0.1.
In the last layer, classification layer,  activation function is employed. 

\subsection{Conclusion}

 In this work, we present a novel RNNs and Attention based model for the drug-drug interactions extraction problem called AW-BLSTMs.
 In this model, a hierarchical layer of BLSTMs is surrounded by two types of attention mechanisms.
 AW-BLSTMs model uses only Words, Part-of-Speech tags as its input features and words embedding representation. This lowest level of external information and syntactic make it easily applicable to different languages and domains which is planned for the future works. Further, focusing on filtering and cleaning, limits the model, and its applicability to real-world scenarios. However, this model has outperformed state-of-the-are DDIs extraction models with F1-score 0.785 by evaluation on the DDIExtraction-2013 corpus.

For the future works, there are different approaches which we would like to investigate. With keeping the generality of model one can apply it on other languages such as the Turkish language. In the other hand, can boost the model by effective features like shortest-dependency-path (SDP) and also, leveraging biomedical ontologies to get a better result in DDI extraction problem. 

\section{Cross-references}
In electronic publications, articles may be internally
hyperlinked. Hyperlinks are generated from proper
cross-references in the article.  For example, the words
\textcolor{black!80}{Fig.~1} will never be more than simple text,
whereas the proper cross-reference \verb+\ref{tiger}+ may be
turned into a hyperlink to the figure itself:
\textcolor{blue}{Fig.~1}.  In the same way,
the words \textcolor{blue}{Ref.~[1]} will fail to turn into a
hyperlink; the proper cross-reference is \verb+\cite{Knuth96}+.
Cross-referencing is possible in \LaTeX{} for sections,
subsections, formulae, figures, tables, and literature
references.

\appendix

\printcredits

\bibliographystyle{cas-model2-names}

\bibliography{cas-refs}


\end{document}